\documentstyle[eqsecnum,aps,epsf]{revtex}
\draft
\begin{document}
\wideabs { 
\title{Improving the false nearest neighbors method with
graphical analysis} \author{T. Aittokallio$^{1,2}$, M. Gyllenberg$^1$,
J. Hietarinta$^3$, T. Kuusela$^{1,3}$, T. Multam\"{a}ki$^3$.}
\address{$^1$Department of Applied Mathematics, University of Turku,
FIN-20014 Turku, Finland\\ $^2$Turku Centre for Computer Science,
University of Turku, FIN-20014 Turku, Finland\\ $^3$Department of
Physics, University of Turku, FIN-20014 Turku, Finland} \date{\today}
\maketitle
\begin{abstract}
We introduce a graphical presentation for the false nearest neighbors
(FNN) method. In the original method only the percentage of false
neighbors is computed without regard to the distribution of
neighboring points in the time-delay coordinates. With this new
presentation it is much easier to distinguish deterministic chaos from
noise. The graphical approach also serves as a tool to determine
better conditions for detecting low dimensional chaos, and to get a
better understanding on the applicability of the FNN method.
\end{abstract}
}

\section{Introduction}
One of the main tasks of time series analysis is to determine from
a given time series the basic properties of the underlying
process, such as nonlinearity, complexity, chaos etc. Among
the most widely used approaches is state space reconstruction by
time delay embedding \cite{Packard80}. After this step has been
taken one can calculate correlation dimensions, various entropy
quantities and estimates for Lyapunov exponents. The crucial
problem is how to select a minimal embedding dimension for the
pseudo phase-space. If the embedding dimension is too small, one
cannot unfold the geometry of the (possible strange) attractor,
and if one uses a too high embedding dimension, most numerical
methods characterizing the basic dynamical properties can produce
unreliable or spurious results.

The false-nearest-neighbors (FNN) algorithm
\cite{Kennel92,Abarbanel93.1,Abarbanel93.2} is one of the tools
that can be used to determine the number of time-delay coordinates
needed to reconstruct the dynamics. In this method one forms a
collection
\begin{equation}
{\bf y}(k)=[x(k),x(k+1),\dots,x(k+d-1)] \label{1}
\end{equation}
of $d$-dimensional vectors for a given time delay (here normalized
to 1), $x(1), x(2),\dots,x(N)$ is a scalar time series. If the
number $d$ of time-delay coordinates in (\ref{1}) is too small,
then two time-delay vectors ${\bf y}(k)$ and  ${\bf y}(l)$ may be
close to each other due to the projection rather than to the
inherent dynamics of the system. When this is the case, points
close to each other may have very different time evolution, and
actually belong to different parts of the underlying attractor.

In order to determine the sufficient number $d$ of time-delay
coordinates one next looks at the nearest neighbor of each vector
(\ref{1}) with respect to the Euclidean metric. We denote the
nearest neighbor of ${\bf y}(k)$ by ${\bf y}(n(k))$. We then
compare the ``$(d+1)$"st coordinates of ${\bf y}(k)$ and ${\bf
y}(n(k))$, e.g., $x(k+d)$ and $x(n(k)+d)$. If the distance $|x(k+d)
- x(n(k)+d)|$ is large the points ${\bf y}(k)$ and ${\bf y}(n(k))$
are close just by projection. They are false nearest neighbors and
they will be pulled apart by increasing the dimension $d$. If the
distances $|x(k+d) - x(n(k)+d)|$ are predominantly small, then
only a small portion of the neighbors are false and $d$ can be
considered a sufficient embedding dimension.

In the FNN algorithm \cite{Kennel92,Abarbanel93.1,Abarbanel93.2} the
neighbor is declared false if
\begin{equation}
\frac{|x(k+d)-x(n(k)+d)|}{\|{\bf y}(k)-{\bf y}(n(k))\|}>R_{tol},
\label{2}
\end{equation}
or if
\begin{equation}
\frac{\|{\bf y}(k)-{\bf y}(n(k))\|^2+[x(k+d)-x(n(k)+d)]^2}{R_A^2}
>A_{tol}^2, \label{3}
\end{equation}
where
\begin{equation}
R_A^2=\frac{1}{N}\sum_{k=1}^N[x(k)-\overline{x}]^2, \label{4}
\end{equation}
and $\overline{x}$ is the mean of all points. The parameter $R_{tol}$
in the first threshold test (\ref{1}) is fixed beforehand, and in most
studies it has been set to $10 - 20$. The second criterion (\ref{3})
was proposed in order to provide correct diagnostics for noise and
usually one takes $A_{tol}\approx 2$. If this test fails, then even
the ($d+1$-dimensional) nearest neighbors themselves are far apart in
the extended $d+1$ dimensional space and should be considered false
neighbors.

Using tests (\ref{2}) and (\ref{3}) one can check all
$d$-dimen\-sional vectors in the data set, and compute the percentage
of false nearest neighbors. By increasing the dimension $d$ this
percentage should drop to zero or to some acceptable small number. In
that case the embedding dimension is large enough to represent the
dynamics.

This method works quite well with noise free data, and the
percentage of false neighbors does not depend on the number of
data points if it is sufficient. However, if data is corrupted
with noise, the percentage of false nearest neighbors for a given
embedding dimension increases as the amount of data is increased,
and therefore a longer time series leads to erroneous false
nearest neighbors as a result of noise corruption rather than of
an incorrect embedding dimension. One possible solution to this
problem is to modify the threshold test (\ref{2}) to account for
additional noise effects. For example, instead of test (\ref{2})
the threshold could be determined by \cite{Rhodes97}
\begin{equation}
\frac{|x(k+d)-x(n(k)+d)|}{\|{\bf y}(k)-{\bf y}(n(k))\|}>R_{tol}
+\frac{2\epsilon R_{tol}\sqrt{d}+2\epsilon} {\|{\bf y}(k)-{\bf
y}(n(k))\|}. \label{5}
\end{equation}
Here the new parameter $\epsilon$ must be chosen properly.
Obviously the optimal value for $\epsilon$ should be determined by
the noise level but unfortunately we have usually very limited
information on the amplitude of the noise in a given time series.

\section{Graphical representation of nearest neighbor
distributions} \label{sec:level1} Without a clear understanding of the
{\em distribution} of neighboring points in the time delay coordinates
the original test (\ref{2}) or the modified test (\ref{5}) cannot
guarantee that we have reached a sufficient embedding dimension, even
if the percentage of false nearest neighbors is low. We have therefore
constructed a simple graphical presentation which simultaneously
displays all essential features. The basic idea is that we show the
distance $R_\Delta = |x(k+d) - x(n(k)+d)|$ as a function of the
original distance $R_d = \|{\bf y}(k) - {\bf y}(n(k))\|$ for all
$d$-dimensional vectors in the data set. The $x$-variable $R_d$ should
be scaled with the normalization coefficient $\sqrt{d}$ in order to
remove unessential changes in the graphs due to changes in the
embedding dimension (see Appendix).

As the first example we have chosen the Henon system
\begin{equation}
X_{n+1}=1-1.4\,X_n^2+Y_n,\quad Y_{n+1}=0.3\,X_n
\label{Henon}
\end{equation}
The parameters of this system were selected from the chaotic region
(the dimension of the attractor is $1.26$), and the total number of
data points is $1000$. In Figure 1 we have plotted $(\tilde
R_d,R_\Delta)$ pairs ($\tilde R_d = R_d/\sqrt{d}$) for all vectors
${\bf y}$. The displayed box size is $0.024\times 0.024$ units. Two
distributions have also been presented in each graph: the $\tilde R_d$
distribution on the bottom part of the graphs, and the radial
distribution plotted on the quarter arc.  The embedding dimension $d$
is scanned from $1$ to $4$, and each set of four graphs is presented
in four different cases where the amplitude of the additional
uniformly distributed (measurement) noise is 0\%, 0.1\%, 1\% and 10\%
of the total amplitude.

According to (\ref{2}) a neighbor is false if it lies above the
straight line going through the origin with slope $R_{tol}$. If we
use the test (\ref{5}) the line has the same slope but there is an
intercept equal to the noise correction term (scaled with
$\sqrt{d}$). Normally we must know the slope a priori but using
these graphs it is not necessary. If there is no noise we clearly
see that with the embedding dimension $>1$ all points lie in the
sector determined by the $x$-axis and a line with slope angle well
below 90 degrees. This important feature can be understood if we
assume that the dynamics is given by
\begin{equation}
x(k+dT)=f(x(k),x(k+1),\dots,x(k+d-1)). \label{9}
\end{equation}
Then we can write
\begin{equation}
\big | x(k+d)-x(l+d)\big | \leq \big\|\nabla f(\xi)\big\|
\big\|{\bf y}(k)-{\bf y}(l)\big\| \label{10}
\end{equation}
for some $\xi$, which implies that
\begin{equation}
\frac{R_\Delta}{R_d}\leq\big\| \nabla f(\xi)\big\|. \label{11}
\end{equation}
Therefore all points in the $(\tilde R_d, R_\Delta)$ plots must
lie under a line which depends on the specific system. The limit
(\ref {11}) is true only when the embedding dimension is
sufficient, and for noise it is never possible. If the time series
includes some additional noise we see its effect as a blurred
border line.

If the embedding dimension is too low the points cumulate close to
the $y$-axis. The radial distribution plot confirms this result.
If $d=1$ the distribution has significant values only with angles
close to $90$ degrees but if $d>1$ the distribution is almost zero
within a distinct range at high angles. The $\tilde R_d$
distribution is high only in the vicinity of zero. A small amount
of noise (0.1\%, the second row from the bottom in Figure 1) does
not change the picture much.

If the level of additional noise is increased to 1\% the points do
not show as well formed pattern. Also the radial distribution is
quite broad but it nevertheless has a clear zero range at high
angles if the embedding dimension is $3$, which can be regarded as
an indication of underlying chaotic (or at least deterministic)
dynamics. The maximum of the $\tilde R_d$ distribution has clearly
shifted towards large values which is typical for pure noise.

In the case of more noisy data (10\% on the top row of Figure 1) the
distribution of points is totally different. Increasing the embedding
dimension does not really change the overall shape of the point
distribution. The radial distribution is fairly even, and the $\tilde
R_d$ distribution is well centered and its maximum shifts toward
higher values when the embedding dimension is increased. (With this
kind of distribution the modified test (\ref{5}) does not really take
noise effects into account.)

In Figure 2 we have presented corresponding graphs for the Lorenz
system
\begin{eqnarray}
\dot X&=&16\,(Y-X)\nonumber\\ \dot
Y&=&X(45.92\,-Z)-Y\label{Lorenz}\\ \dot Z&=&XY-4\,Z \nonumber
\end{eqnarray}
using $10000$ data points and the sampling delay of $0.05$. For these
parameter values the dimension of the attractor is $2.07$.  Here we
observe similar kind of behavior for various distributions as in the
case of the Henon system. Since the true dimension of the attractor is
greater than $2$, a clearly bounded sector pattern of points can only
be seen in the graphs with embedding dimension $\geq 3$.  For $d=2$
most of the points lie under a line with slope under $90$ degrees
which is also reflected in the noticeable maximum of the radial
distribution, and since there is only a small portion of points between
this maximum and the $y$-axis we can estimate that the true dimension
of the attractor is not much greater than $2$.

The effect of even a small amount of noise can be clearly seen in
Figure 2. Already with 1\% of noise the sector pattern has changed to
a vertical one. This is shown clearly in the regression lines
(corresponding to the first principal component of the points $(\tilde
R_d, R_\Delta)$) plotted in Figure 2. In the two bottom rows the
regression lines have a slope well below $90$ degrees, and this can be
taken as evidence of deterministic dynamics. For the two top rows the
regression line is almost vertical (see also Figure 3) indicating
noise contamination.  Furthermore we see that the $\tilde R_d$
distribution shows approximately Gaussian shape, which spreads out and
moves further and further away from the origin as the noise level or
embedding dimension increases. The radial distribution, on the other
hand, moves closer to the $90$-degrees line as noise contamination
increases, which means that the height/width ratio of the point
distribution increases, and therefore that it is more and more
difficult to predict the next point.

In the standard procedure noise effect are taken into account by the
condition (\ref 3), which means that points outside a circle of radius
$A_{tol}R_A$ are counted false (actually it is an ellipse, due to the
scaling of $\tilde R_d$.) For Figures 2 and 4 this radius is 500 times
the box size (and for figures 1 and 5 the factor is about 20).
Although the boundary is quite far away one can imagine that higher
levels of noise and higher embedding dimensions both increase the
number of false neighbours, as has been reported
\cite{Abarbanel93.1,Abarbanel93.2}.

If the total number of data points of the preceeding system is
decreased to $1000$ the graphs are not so simple to interpret (Figure
4). There is no significant difference between graphs with embedding
dimension 2 and 3. As usual, reliable estimation of the underlying
dynamical dimension requires a sufficient number of data points.
However, by using this graphical representation we can nevertheless
make a rough estimate on dimension even when only relatively few data
points are available.

As a final example we have analyzed the Mackey-Glass system
\begin{equation}
\dot X=\frac{0.2\,X(t+31.8)}{1+[X(t+31.8)]^{10}}-0.1\,X(t)
\label{Mackey}
\end{equation}
using the sampling delay of 2. As the dimension of the attractor
with these parameter values is about $3.6$, the embedding
dimension must be at least 4. This can be seen in Figure 5: only
in rightmost graph there is a clear sector type of pattern, and
the radial distribution is zero over a nonzero range of angles
near $90$ degrees.

\section{Conclusions}
We have presented a graphical method to analyze time series in order
to estimate the sufficient embedding dimension and the portion of
additional noise. This tool consists of a $(\tilde R_d, R_\Delta)$
plot augmented with two distributions. Furthermore, the slope of the
regression line of points in the $(\tilde R_d, R_\Delta)$ graphs can
be used to recognize noise in deterministic systems.

The advantage of the present method is that even small amount of noise
contamination can be distinguished from deterministic chaos. This also
means that we now see how the problem of determining the correct
embedding dimension becomes more difficult with even a small amount of
noise, and that for a deterministic system where the proportion of
noise is substantial one should use the conditions (\ref{2}) or
(\ref{5}) with great caution. If the FNN algorithm is used to estimate
the embedding dimension, our presentation should be used in parallel
in order to get relevant and reliable results.

To summarize our method we present a list of guidelines on how to
distinguish a deterministic time series from sources with
noise:

\vskip 0.2cm
\noindent The time series is produced by a
deterministic system if:
\begin{enumerate}
\item the points in the $(\tilde R_d, R_\Delta)$ plot form a clear
sector pattern with a zero radial distribution over a distinct
range below $90$ degrees,
\item the $R_d$ distribution is centered close to zero,
\item the slope of the regression line is well below 90
degrees.
\end{enumerate}
The noise level in the time series is substantial if:
\begin{enumerate}
\item the radial distribution is spread out over the whole range
from 0 to 90 degrees,
\item the $R_d$ distribution has a clear maximum far away from zero,
\item the slope of the regression line is close to $90$ degrees.
\end{enumerate}

\appendix
\section*{}
Let $f$ be a function which has been sampled very densely. Then we
can assume that the nearest neighbor of the $d$-dimensional vector
is the vector that starts at the next (or previous) sample point
\begin{equation}
\bigl(f(t_0+\delta),f(t_0+2\delta),f(t_0+3\delta),\dots,f(t_0+d\delta)
\bigr).
\label{6}
\end{equation}
The distance between these two points is therefore
\begin{eqnarray}
R_d&&\,=\sqrt{\sum_{i=1}^d\bigl(f(t_0+i\delta)-f(t_0+(i-1)\delta)
\bigr)^2}\nonumber\\ &&\approx \sqrt{\sum_{i=1}^d\delta^2
f'(t_0+i\delta)^2} \approx \delta\sqrt{d}|f'(t_0)|, \label{7}
\end{eqnarray}
where we have assumed that the function $f$ changes relatively
slowly (or that it is linear). The distance between the targets is
\begin{equation}
R_\Delta =\big |f(t_0+\delta +1)-f(t_0+\delta)\big |
\approx\delta\big |f'(t_0)\big |, \label{8}
\end{equation}
and by combining the results (\ref{7}) and (\ref{8}) we conclude
that the ratio of $R_\Delta/R_d$ is $1/\sqrt{d}$, and therefore is
it reasonable in all cases to normalize this ratio with
$\sqrt{d}$.

\widetext
\newpage
\ 
\newpage

\begin{figure*}
\begin{center}
\epsfxsize=17.5cm
\epsffile{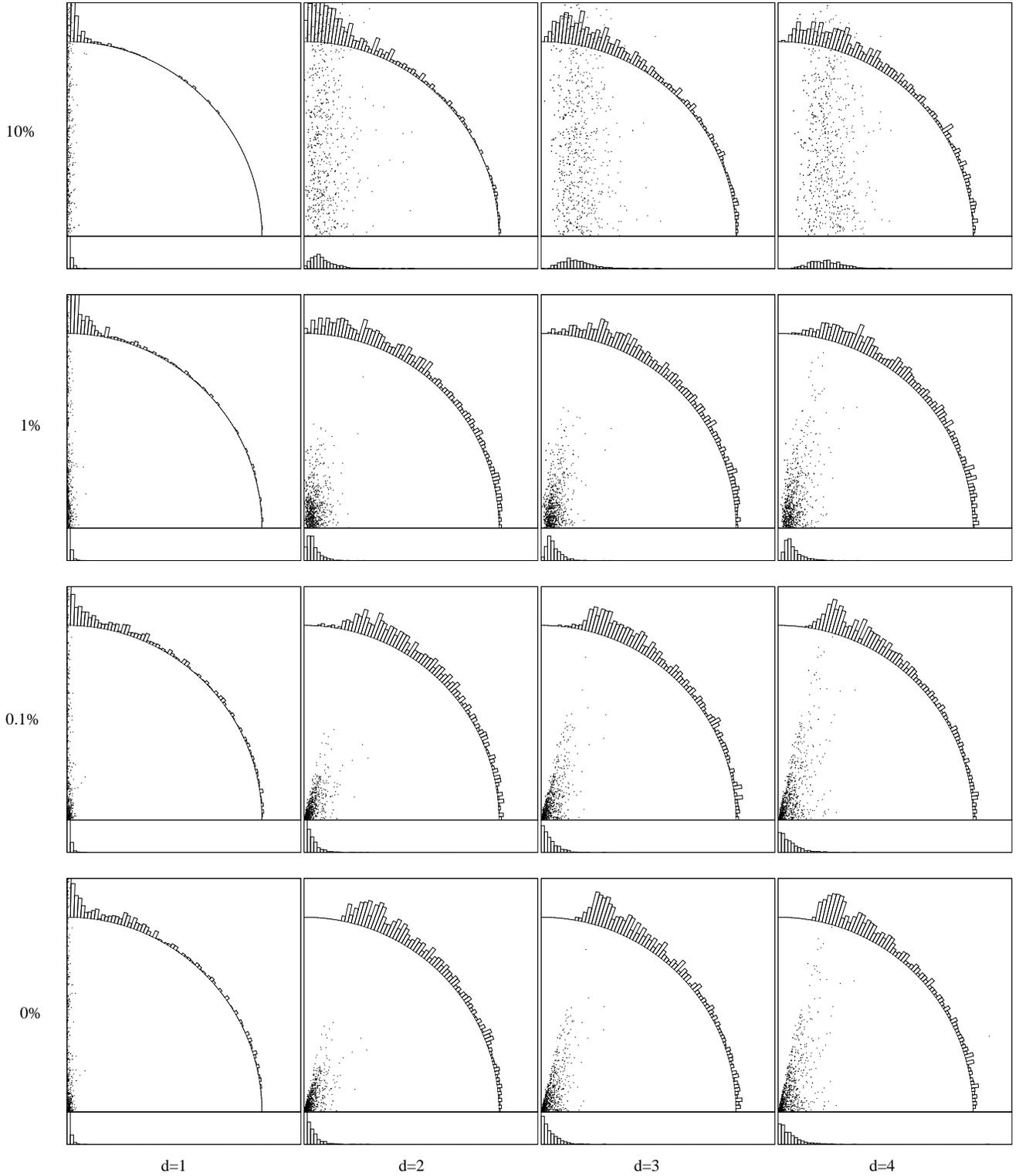}
\caption{ The target distance $R_\Delta$ as a function of the nearest
neighbor distance $\tilde R_d$ for the Henon system (the dimension
of the attractor is $1.26$). The total number of data points is
$1000$. The rows correspond to indicated noise levels, the columns
to indicated embedding dimensions.}
\end{center}
\end{figure*}

\newpage
\
\newpage

\begin{figure*}
\begin{center}
\epsfxsize=17.5cm
\epsffile{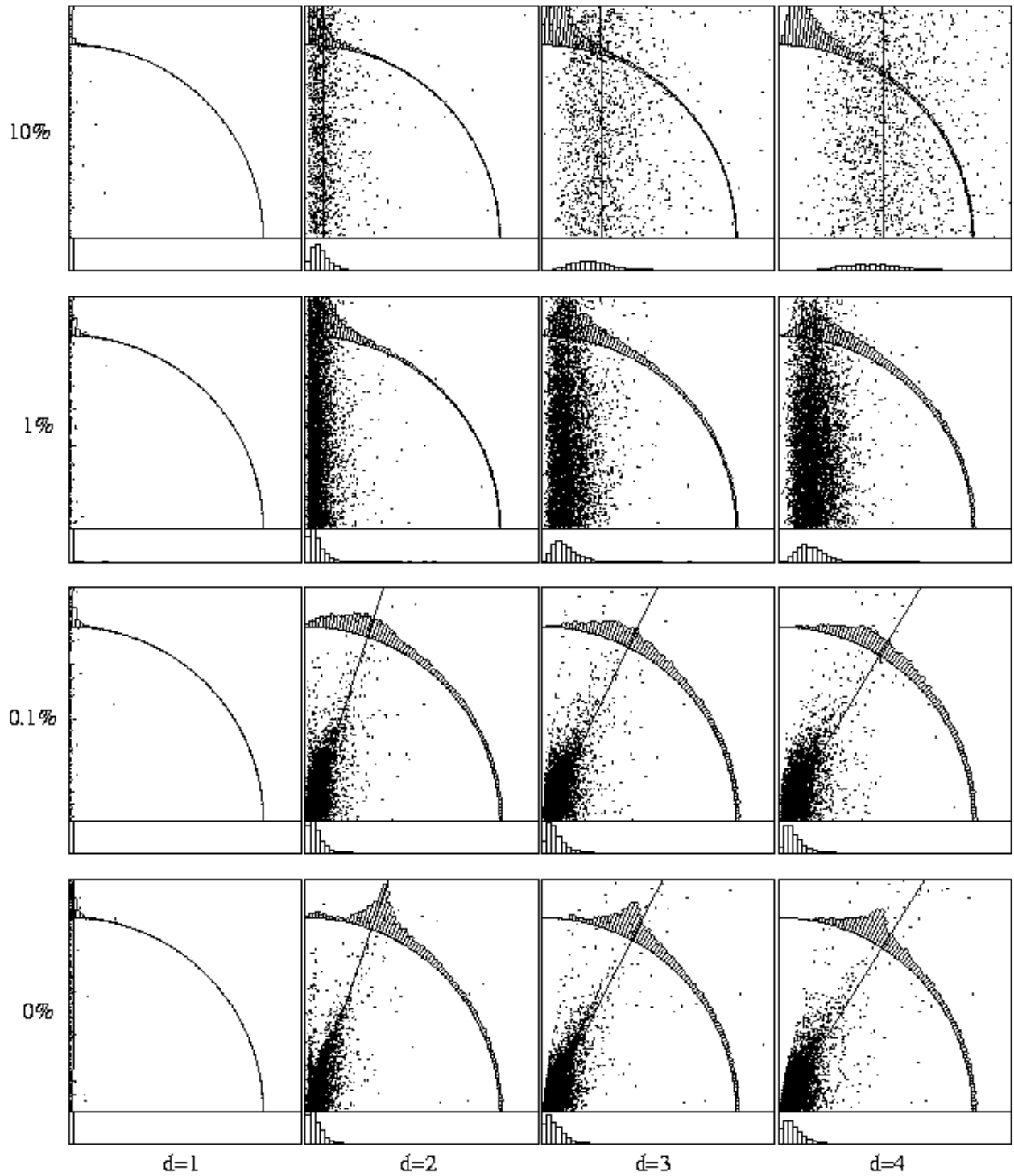}
\caption{Same as in Figure 1 but for the Lorenz system (the dimension
of the attractor is 2.06). The total number of data points is $10000$.
The regression lines are also plotted on each graph. (We apologize for
the low resolution of this figure, the original PostScript file was
too big to store at this archive, but it is available at {\tt
http://www.utu.fi/~hietarin/chaos/fnn.html}.)}
\end{center}
\end{figure*}

\newpage
\
\newpage

\narrowtext
\begin{figure*}
\begin{center}
\epsfxsize=6cm
\epsffile{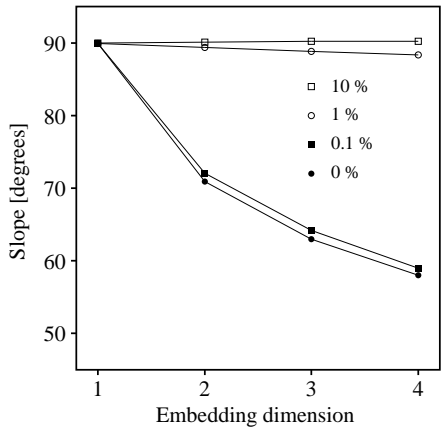}
\caption{The slope of the regression line as a function of the
embedding dimension for different percentage of noise taken from
the graphs in Figure 2.}
\end{center}
\end{figure*}
\widetext

\begin{figure*}
\epsfxsize=17.5cm
\epsffile{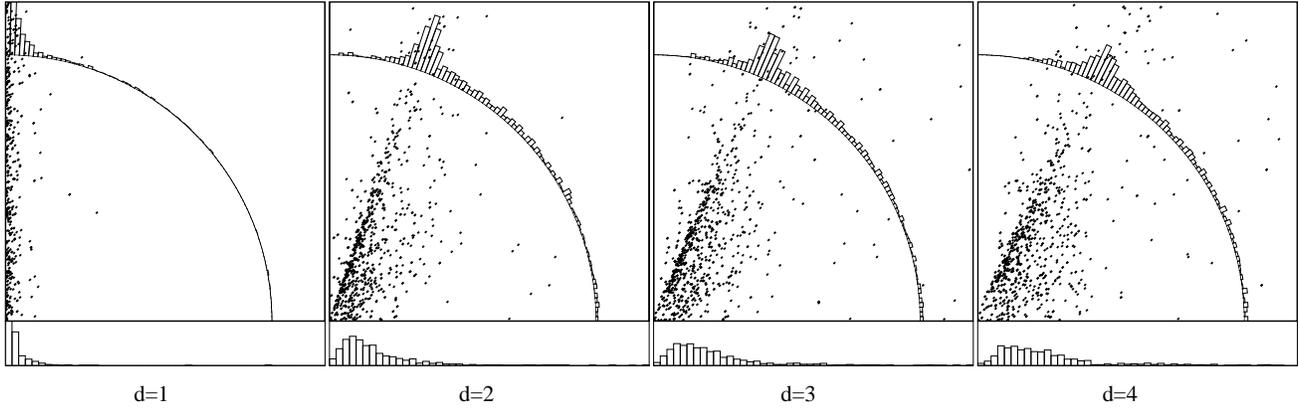}
\caption{The same graphs as in the bottom row of Figure 2 but
the total number of data points is only $1000$.}
\end{figure*}

\begin{figure*}
\begin{center}
\epsfxsize=17.5cm
\epsffile{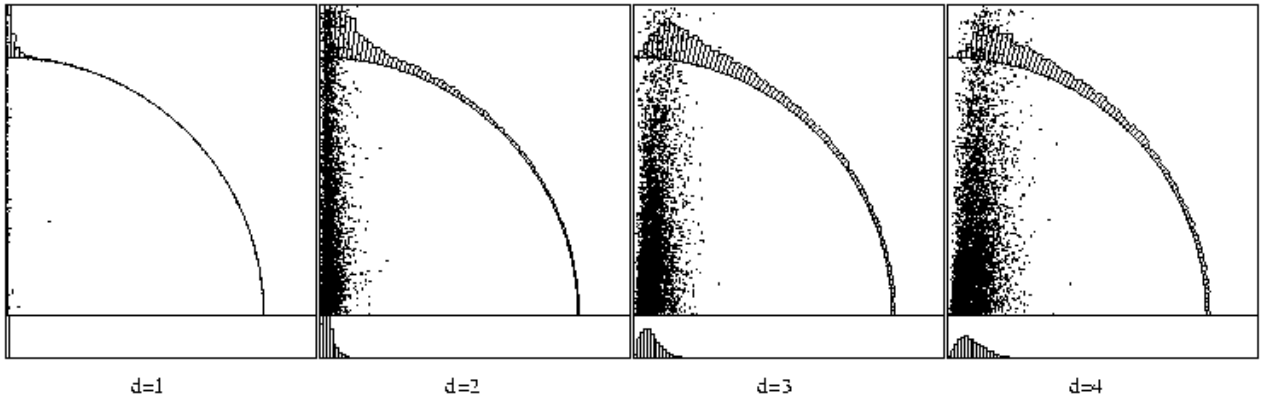}
\caption{The target distance R$_\Delta$ as a function of the nearest
neighbor distance $\tilde R_d$ for the Mackey-Glass system (the
dimension of the attractor is $\sim 3.6$). The total number of data
points is $10000$.  (We apologize for the low resolution of this
figure, the original PostScript file was too big to store at this
archive, but it is available at {\tt
http://www.utu.fi/~hietarin/chaos/fnn.html}.)}
\end{center}
\end{figure*}

\end{document}